\documentclass[pra,twocolumn,tightenlines,nofootinbib,superscriptaddress,10pt,aps]{revtex4-2}
\usepackage{bm,dcolumn,amsmath,graphicx,amsfonts,amssymb}

\usepackage[version=4]{mhchem} 
\usepackage{siunitx} 
\usepackage{physics} 

\usepackage{hyperref,xcolor}
\definecolor{cite}{rgb}{0.,0.,0.9}
\hypersetup{colorlinks,linkcolor={cite},citecolor={cite},urlcolor={cite}}

\newcommand{\rme}[3]{
\langle{#1}\|{#2}\|{#3}\rangle
}

\newcommand{\smallspace}{\rule{0pt}{3.3ex}}

\begin{document}

\title{Theoretical characterization of the barium II and radium II ions}

\date{September 2025} 

\author{Robin~B.~Cserveny}
\email[]{r.cserveny@uqconnect.edu.au}
\affiliation{Faculty of Physics, University of Vienna, 1090 Vienna, Austria}
\affiliation{School of Mathematics and Physics, The University of Queensland, Brisbane QLD 4072, Australia}

\author{Benjamin~M.~Roberts}\email[]{b.roberts@uq.edu.au}
\affiliation{School of Mathematics and Physics, The University of Queensland, Brisbane QLD 4072, Australia}

\begin{abstract}
\noindent
Motivated by recent experimental advances, including the ongoing development of an optical atomic clock in singly ionized radium, we perform a detailed theoretical characterization of \ce{Ra+} and its lighter analogue, \ce{Ba+}.
Both ions are of interest for precision studies, including for atomic parity violation and searches for new physics beyond the standard model.
Using the all-orders correlation potential method, including Breit and radiative quantum electrodynamics corrections, we perform high-accuracy calculations of electric-dipole (E1), electric-quadrupole (E2), and magnetic-dipole (M1) transition matrix elements between the low-lying $s$, $p$, and $d$ states of these ions, as well as the excited-state lifetimes, polarizabilities, magic wavelengths, and magnetic dipole (A) hyperfine structure constants. 
By combining lifetime measurements with accurate theoretical ratios, we extract high-precision determinations of the $s-d_{1/2,\,3/2}$ E2 matrix elements.
By combining hyperfine measurements with atomic theory, we extract parameters of the nuclear magnetization distribution (the Bohr-Weisskopf effect) for $^{135,\,137}$Ba and $^{223,\,225}$Ra.
These results provide theoretical input for ongoing and future experimental programs in fundamental physics and precision metrology.
\end{abstract}

\maketitle

\section{Introduction}

Recent advances in high-precision spectroscopy of singly-ionized barium and radium have generated renewed interest in these systems for fundamental studies and optical clock development.
In \ce{Ba+}, precise measurements of $p$ state lifetimes and branching fractions~\cite{Arnold2019,Zhang2020a}, and magic wavelengths~\cite{Chanu2020} have been reported.
For \ce{Ra+}, recent progress includes precise determinations of $d$ state lifetimes~\cite{Li-arXiv:2501.18866}, $p$ state branching fractions and lifetimes~\cite{Fan:2019eze,Fan:2019,Fan2022}, transition frequencies~\cite{Kofford:2025,Holliman2020}, and precise determination of the hyperfine structure of \ce{^{225}Ra+}~\cite{Ready:2024xxx}.

At the same time, recent theoretical work has highlighted the importance of higher-order correlations~\cite{Porsev:2021fhb}, radiative quantum electrodynamics (QED)~\cite{Fairhall:2022plc}, and nuclear structure~\cite{Ginges:2017rmj} effects in these ions.
Particularly high theoretical accuracy was recently demonstrated for electric dipole matrix elements~\cite{Roberts:2022lda} in these and other monovalent atoms and ions.

These efforts support ongoing work towards high-accuracy optical atomic clocks based on \ce{Ra+}~\cite{Versolato:2011-83.043829, Holliman:2022-128.033202, Holliman-arXiv:2211.05313} and \ce{Ba+}~\cite{Fujisaki:2025.94.074304} ions, and motivate continued theoretical development to match the level of precision now achievable.
Optical atomic clocks, with their unparalleled stability and accuracy, are powerful tools for probing fundamental physics~\cite{Safronova:2017xyt}.
They enable high-precision measurements that can constrain possible variations in fundamental constants, such as the fine-structure constant~\cite{Flambaum:2008kr}.
In addition, these systems serve as sensitive probes of exotic physics, including through studies of atomic parity violation (APV)~\cite{Roberts:2014bka}, searches for electric dipole moments~\cite{Ginges:2003qt}, and searches for dark matter~\cite{Filzinger:2023qqh}.

Progress has been made towards measuring APV in \ce{Ra+}, presenting a promising avenue for precision tests of the Standard Model at low energies~\cite{NunezPortela:2014-114, Versolato:2010man, Williams:2013-88.012515, Fan:2019eze}.
The heavy atomic nucleus of radium ($Z=88$) amplifies APV effects, which scale faster than $Z^3$~\cite{Bouchiat_1974}, making \ce{Ra+} an excellent candidate for such studies~\cite{Dzuba:2001zz, Roberts:2013cbw, Wansbeek:2008yma, Sahoo:2006-96.163003, Pal2009, Li:2021.274.107877, Li:2021ihn}.
Such measurements could, in theory, improve upon the current measurements for \ce{Cs}~\cite{Wood:1997zq, Dzuba:2002kx, Porsev:2009pr, Dzuba_2012}.
Additionally, the lowest $s - d$ transitions in \ce{Ra+} are known to be very sensitive to variations in the fine-structure constant~\cite{Dzuba:1999zv}, and transitions in the \ce{^{133}Ba+} ion are of interest for quantum information applications~\cite{Hucul.119.100501:2017}.

In this work, we present high-accuracy theoretical calculations of electric dipole (E1), electric quadrupole (E2), and magnetic dipole (M1) transition matrix elements, excited-state lifetimes, branching fractions, polarizabilities, magic wavelengths, and hyperfine structure constants for \ce{Ba+} and \ce{Ra+}.
We use the all-orders correlation potential method, including Breit and QED corrections.
By combining experimental measurements with accurate theory, we extract improved values for E2 matrix elements and Bohr-Weisskopf corrections (parameter for the finite nuclear magnetization distribution), enabling refined tests of atomic structure and nuclear effects.
These results provide essential theoretical input for ongoing efforts in atomic parity violation, clock development, and searches for new physics beyond the Standard Model.

\section{Theory}

For the calculations, we use an all-orders correlation potential method based on the Feynman diagram technique developed in Refs.~\cite{DzubaCPM1988pla, DzubaCPM1989plaEn}, which provides high accuracy for these systems, with relatively minimal computational costs.
The specific implementation, which we refer to as atomic many-body perturbation theory in the screened Coulomb interaction (\textsc{ampsci}), is described in detail in Ref.~\cite{Roberts:2022lda}.
Here, we provide a brief overview of the method.

The calculations begin with the relativistic Hartree-Fock (RHF) method in the $V^{N-1}$ approximation, where the wavefunction for the valence electron is found in the frozen potential of the $N$\,$-$\,1 core electrons.
To account for core-valence correlations, the correlation potential, $\Sigma$, is added to the RHF equation for the valence states:
\begin{equation}
    (h_{\rm HF} + \hat \Sigma_\varepsilon - \varepsilon)\psi = 0,
\end{equation}
where $h_{\rm HF}$ is the regular single-particle Hartree-Fock Hamiltonian, and $\varepsilon$ is the single-particle energy.
The resulting orbitals are known as Brueckner orbitals.
The most significant factor affecting the accuracy is the order to which $\Sigma$ is calculated.
It may be calculated to lowest (second) order in the Coulomb interaction using the standard Goldstone approach (see, e.g., Ref.~\cite{Lindgren}), which we refer to as $\Sigma^{(2)}$.
For accurate calculations, higher-order effects must be included.
In the all-orders correlation potential method, three important classes of diagrams are included to all orders:\ (i) screening of the core-valence Coulomb interaction by the core electrons~\cite{DzubaCPM1988pla}, (ii) the hole-particle interaction~\cite{DzubaCPM1989plaEn}, and (iii) the chaining of the correlation potential, which is included automatically by solving the Brueckner equation.
We refer to the all-orders correlation potential as $\Sigma^{(\infty)}$.

We account for the most important radiative quantum electrodynamics (QED) effects using the radiative potential method~\cite{Flambaum:2005ni, Ginges2016}, which includes the dominating vacuum polarization and approximate self-energy corrections into atomic energies and wavefunctions.
The corrections to certain matrix elements can be included this way, and as demonstrated in Ref.~\cite{Fairhall:2022plc}, improve the accuracy for electric dipole transitions.
We note that certain effects, in particular vertex corrections, are excluded in this method.
For the dipole matrix elements, the vertex corrections are negligible~\cite{Flambaum:2005ni, Fairhall:2022plc}; the same is expected for the E2 matrix elements.
For the hyperfine interaction, on the other hand, the vertex corrections are important (see, e.g., Ref.~\cite{Flambaum:2005ni, Roberts:2021esp}).
As such, for hyperfine constants, QED corrections are instead estimated by rescaling calculations from Ref.~\cite{Ginges:2017fle}.

We include the leading-order relativistic corrections to the electron-electron Coulomb interaction via the Breit approximation (see, e.g., Ref.~\cite{Mann1971}).
The interplay between Breit and correlation effects are crucial~\cite{Derevianko:2000dt}, and we calculate the Breit correction at the level of second-order many-body perturbation theory.
We also include the Breit correction to the correlation potential, though this is negligible in all cases.

Finally, we include small semi-empirical correction by rescaling the correlation potential to reproduce the experimental energies:
\begin{equation}\label{eq:scale}
    \Sigma\to\lambda\Sigma,
\end{equation}
where $\lambda\approx1$, and is allowed to vary separately for each valence state.
For example, for the lowest valence states of \ce{Ba+}, we have $\lambda=0.99$ for $s$-states, 0.98 for $p$-states, and 0.91 for $d$-states.
This scaling procedure estimates the contributions from missed higher-order diagrams, and leads to improved wavefunctions and matrix elements.
Furthermore, comparing the results using the scaled and un-scaled correlation potentials at the second- and all-order levels gives an excellent handle on the theoretical uncertainty stemming from missed correlation effects~\cite{Roberts:2022lda}.

\subsection{External fields}

The interaction with external fields leads to a modification of the wavefunctions of the core electrons: $\psi_c\to\psi_c + \delta\psi_c$, which leads to a correction to the Hartree-Fock potential, $\delta V$.
This, in turn, gives a correction to matrix elements known as core polarization:
\begin{equation}\label{eq:matel}
    \langle\psi_v|h_{\rm hfs} + \delta V|\psi_w\rangle.
\end{equation}
It is found to first-order in the external field via the time-dependent Hartree-Fock (TDHF) method~\cite{Dzuba1984:CsFr}, equivalent to the diagrammatic random phase approximation (RPA)~\cite{Johnson:1980-21.409}.
The Breit interaction also modifies the TDHF equations as well as the resulting $\delta V$ correction.

The main correlation effects are included into matrix element via use of Brueckner orbitals in Eq.~\eqref{eq:matel}.
We also account for the structure radiation (external field correction to the correlation potential), and the renormalization (shift in the normalization due to the correlation corrections)~\cite{Dzuba1987,Johnson1987}.

The matrix elements for the non-relativistically forbidden M1 transitions are very small.
In numerical calculations, this smallness is manifest through the near-total cancellation between two order 1 terms.
As such, even very small numerical errors stemming from inexact orthogonalization can lead to significant errors in the final values~\cite{Safronova2017}.
In such cases, we perform our calculations also using a third-order method based entirely on a basis expansion~\cite{Johnson1987}, where the orthogonality can be guaranteed to a high level of accuracy.
In these cases, we do not include the Breit or QED corrections, since they enter well below the assumed correlation uncertainty.

\begin{table*}
\centering
\caption{Calculated {\sl ab initio} removal energies (cm$^{-1}$) for the lowest few valence states of \ce{Ba+} and \ce{Ra+}, and comparison with experiment. The calculations are shown at the relativistic Hartree-Fock (RHF) and all-orders correlation potential levels; the `Final' column further includes the Breit and QED corrections.}
\label{tab:en}
\begin{ruledtabular}
    \begin{tabular}{lD{.}{.}{5.1}D{.}{.}{5.1}D{.}{.}{5.1}D{.}{.}{5.1}D{.}{.}{2.3}
    |lD{.}{.}{5.1}D{.}{.}{5.1}D{.}{.}{5.1}D{.}{.}{5.1}D{.}{.}{2.3}}
     \ce{Ba+} & \multicolumn{1}{c}{RHF} & \multicolumn{1}{c}{$\Sigma^{(\infty)}$} & \multicolumn{1}{c}{Final} & \multicolumn{1}{c}{Expt. \cite{NIST-ASD:2024}} & \multicolumn{1}{c|}{$\Delta$(\%)} & 
     \ce{Ra+} & \multicolumn{1}{c}{RHF} & \multicolumn{1}{c}{$\Sigma^{(\infty)}$} & \multicolumn{1}{c}{Final} & \multicolumn{1}{c}{Expt. \cite{NIST-ASD:2024}} & \multicolumn{1}{c}{$\Delta$(\%)} \\ 
     \hline
        6$s_{1/2}$ & 75339.7 & 80774.3 & 80722.6 & 80686.3 & 0.04   & 7$s_{1/2}$ & 75897.5 & 81999.5 & 81899.7 & 81842.5 & 0.07   \\
        7$s_{1/2}$ & 36851.7 & 38324.8 & 38308.0 & 38331.1 & -0.06  & 8$s_{1/2}$ & 36859.9 & 38440.6 & 38409.1 & 38437.5 & -0.07  \\
        8$s_{1/2}$ & 22023.3 & 22656.5 & 22649.1 & 22661.1 & -0.05  & 9$s_{1/2}$ & 22004.3 & 22677.5 & 22663.4 & 22677.2 & -0.06  \\
        6$p_{1/2}$ & 57265.5 & 60513.3 & 60486.7 & 60424.7 & 0.10   & 7$p_{1/2}$ & 56878.2 & 60634.4 & 60583.7 & 60491.2 & 0.15   \\
        7$p_{1/2}$ & 30240.1 & 31312.3 & 31302.3 & 31296.5 & 0.02   & 8$p_{1/2}$ & 30052.9 & 31254.6 & 31235.8 & 31236.2 & -0.001 \\
        8$p_{1/2}$ & 18848.0 & 19350.8 & 19346.2 & 19346.8 & -0.003 & 9$p_{1/2}$ & 18748.1 & 19291.3 & 19283.1 &         &        \\
        6$p_{3/2}$ & 55873.3 & 58794.9 & 58786.0 & 58733.9 & 0.09   & 7$p_{3/2}$ & 52905.7 & 55698.5 & 55685.7 & 55633.6 & 0.09   \\
        7$p_{3/2}$ & 29699.0 & 30686.1 & 30682.3 & 30675.0 & 0.02   & 8$p_{3/2}$ & 28502.2 & 29459.9 & 29454.3 & 29450.5 & 0.01   \\
        8$p_{3/2}$ & 18579.6 & 19046.8 & 19045.2 & 19044.3 & 0.005  & 9$p_{3/2}$ & 17975.2 & 18410.2 & 18408.7 & 18432.1 & -0.13  \\
        5$d_{3/2}$ & 68139.0 & 76444.5 & 76522.0 & 75812.4 & 0.94   & 6$d_{3/2}$ & 62355.9 & 70192.0 & 70301.3 & 69758.2 & 0.78   \\
        6$d_{3/2}$ & 33266.3 & 34741.3 & 34752.5 & 34736.8 & 0.05   & 7$d_{3/2}$ & 31575.0 & 33091.3 & 33109.1 & 33098.5 & 0.03   \\
        5$d_{5/2}$ & 67664.8 & 75612.0 & 75710.1 & 75011.5 & 0.93   & 6$d_{5/2}$ & 61592.5 & 68501.1 & 68627.9 & 68099.5 & 0.78   \\
        6$d_{5/2}$ & 33093.3 & 34533.6 & 34549.8 & 34531.5 & 0.05   & 7$d_{5/2}$ & 31203.6 & 32596.4 & 32620.0 & 32602.1 & 0.06   \\
    \end{tabular}%
\end{ruledtabular}
\end{table*}

\subsection{Uncertainty Estimation}

We estimate the uncertainty in our calculations by comparing our results at different orders of approximation.
Specificity, we use the method similar to that outlines in Ref.~\cite{Roberts:2022lda}, where it was demonstrated to be robust.
To estimate the uncertainty in the quantity $M$ due to omitted correlation effects, we consider the differences
\[
    |M({\lambda \Sigma^{(\infty)}}) - M({\lambda \Sigma^{(2)}})| 
    \quad \text{and} \quad 
    |M({\lambda \Sigma^{(\infty)}}) - M({\Sigma^{(\infty)}})|.
\]
In most cases, we take the uncertainty associated with the correlation potential (the ``Brueckner'' uncertainty) to be the maximum of these two values.  
For matrix elements involving $d$-states---where the uncertainty in our method is expected to be larger---we instead take the sum of these differences as the Brueckner uncertainty.
Importantly, this approach ensures that our quoted uncertainties exceed the difference between the scaled and unscaled {\sl ab initio} results.
To include non-Brueckner type effects, we take half the Breit and QED contributions, and 30\% of the structure radiation and normalization corrections.

Finally, for the hyperfine constants, there is an additional source of uncertainty arising from nuclear structure -- specifically, from the spatial distribution of the nuclear magnetic and electric moments.
In particular, for the magnetic dipole hyperfine constants $A$, the so-called Bohr-Weisskopf effect, has been shown to be significant at the current level of theoretical accuracy~\cite{Ginges:2017rmj}.
The details of the finite nuclear distributions are discussed below.

\section{Energies and transitions}

The calculated binding energies for several of the low-lying valence states for each ion are presented in Table~\ref{tab:en}.
The agreement with experiment is excellent for nearly all states, with discrepancies at or significantly below the 0.1\% level.
The exception is the lowest $d$ states of both ions.
The discrepancy for the $d$ states is due to missed high-order correlation diagrams (in particular, the so-called ladder diagrams~\cite{DzubaLadder2008}), which are significant for these states due to their low principle quantum number, and higher overlap with the core.
It has been shown (see, e.g., Ref.~\cite{Roberts:2022lda}) that the scaling procedure~\eqref{eq:scale} can well account for the resulting missed corrections to matrix elements.
This is also demonstrated by the excellent agreement between theory and experiment for the $p$ and $d$-state lifetimes calculated in this work, which are determined mainly by $p-d$ E1 and $s-d$ E2 matrix elements (see below).

Not also that the deviations for both ions are qualitatively very similar.
This is easily understood; their electronic structure is very similar.
Compared to barium, radium has a filled $4f^{14}$-shell in the core, which is the most significant qualitative difference.
However, this shell is deep enough in the core so as not to impact the relative correlation effects significantly.
As such, theoretical uncertainties are expected to behave similarly in both systems, and it is therefore useful to treat one as a benchmark for the other when experimental data is only available for one of the ions.


A summary of the calculated reduced matrix elements for the electric dipole (E1), magnetic dipole (M1), and electric quadrupole (E2) transitions between the lowest few $s$, $p$, and $d$ states of \ce{Ba+} and \ce{Ra+} are presented in Table~\ref{tab:matel-summary}.
The E1 matrix elements were calculated recently by one of us using the same method~\cite{Roberts:2022lda}, and are reproduced in Table~\ref{tab:matel-summary} for convenience.
Tables of transitions between all of the states in Table~\ref{tab:en}, including a breakdown of the calculations at different levels of approximation, are presented in the appendix (Table~\ref{tab:E1_BaRa}).
As discussed in Ref.~\cite{Roberts:2022lda}, the accuracy for the E1 matrix elements is excellent.

\begin{table}
    \centering
    \caption{Summary of reduced matrix elements between the lowest few states of \ce{Ba+} and \ce{Ra+};
    transitions between all states in Table~\ref{tab:en} are presented in the appendix. Numbers in square brackets denote powers of 10.}
    \label{tab:matel-summary}
    \begin{ruledtabular}
    \begin{tabular}{lll|llll}
    & &\multicolumn{1}{c}{$E1$ ($ea_0$)~\cite{Roberts:2022lda}} & & & \multicolumn{1}{c}{$M1$ ($\mu_B$)} & \multicolumn{1}{c}{$E2$ ($ea_0^2$)} \\
    \hline
    \multicolumn{7}{c}{\ce{Ba+}}\\
    \smallspace
    6$s$ & 6$p_{1/2}$ & 3.3214(43) & 6$s$ & 7$s$ & -7.2(36)\,[-5] &  \\
    6$s$ & 6$p_{3/2}$ & 4.6886(60) & 6$s$ & 5$d_{3/2}$ & 16.2(33)\,[-5] & 12.55(13) \\
    7$s$ & 6$p_{1/2}$ & 2.478(14) & 6$s$ & 5$d_{5/2}$ &  & 15.71(16) \\
    7$s$ & 6$p_{3/2}$ & 3.860(19) & 7$s$ & 5$d_{3/2}$ & 8(2)\,[-5] & 4.66(22) \\
    5$d_{3/2}$ & 6$p_{1/2}$ & 3.036(31) & 7$s$ & 5$d_{5/2}$ &  & 6.06(28) \\
    5$d_{3/2}$ & 6$p_{3/2}$ & 1.327(14) & 6$p_{1/2}$ & 6$p_{3/2}$ & 1.15231(6) & 28.27(8) \\
    5$d_{5/2}$ & 6$p_{3/2}$ & 4.087(44) & 5$d_{3/2}$ & 5$d_{5/2}$ & 1.54928(21) & 6.66(10) \\
    \hline
    \\
    \multicolumn{7}{c}{\ce{Ra+}}\\
    \smallspace
    7$s$       & 7$p_{1/2}$ & 3.2357(48) & 7$s$       & 8$s$       & 108(40) [-5] &           \\
    7$s$       & 7$p_{3/2}$ & 4.4927(66) & 7$s$       & 6$d_{3/2}$ & 144(40) [-5] & 14.65(12) \\
    8$s$       & 7$p_{1/2}$ & 2.516(17)  & 7$s$       & 6$d_{5/2}$ &              & 18.75(15) \\
    8$s$       & 7$p_{3/2}$ & 4.637(20)  & 8$s$       & 6$d_{3/2}$ & 58(23) [-5]  & 7.56(26)  \\
    6$d_{3/2}$ & 7$p_{1/2}$ & 3.536(26)  & 8$s$       & 6$d_{5/2}$ &              & 10.46(35) \\
    6$d_{3/2}$ & 7$p_{3/2}$ & 1.501(13)  & 7$p_{1/2}$ & 7$p_{3/2}$ & 1.1340(5)    & 29.79(10) \\
    6$d_{5/2}$ & 7$p_{3/2}$ & 4.789(42)  & 6$d_{3/2}$ & 6$d_{5/2}$ & 1.5511(18)   & 8.55(11)  \\
    \end{tabular}%
    \end{ruledtabular}
\end{table}

The calculations for the M1 matrix elements that are forbidden at the non-relativistic level are known to be numerically unstable.
In practice, this instability arises due to very large cancellations between the lowest-order correlation correction to the wavefunctions (``Brueckner'' corrections), and the structure radiation correction.
The cancellations are very sensitive to orthogonality properties of the orbitals, as discussed in detail in Ref.~\cite{Safronova2017}.

As such, we use the numerically stable third-order method as outlined in Ref.~\cite{Blundell:1987hmf} for the calculations for these M1 transitions.
The correspondingly larger uncertainty is conservatively taken to be half the difference between the calculations at the RPA and third-order levels.
The impact of basis truncation errors has also been checked, and is significantly below this assumed uncertainty.
We account for the frequency-dependence of the relativistic M1 operator in all cases; while this makes a significant impact at the Hartree-Fock level, the frequency-dependent contribution becomes very small after RPA corrections are included.
We do not include the Breit or QED corrections here, since their impacts are significantly below the level of uncertainty.

\begin{table}
\caption{Branching fractions for decays from the lowest few states of \ce{Ba+}. 
The $5d_{3/2}$ state decays only to $6s$.
Comparison with experiment~\cite{Arnold2019,Zhang2020a} (in italics) shows excellent agreement.}
\label{tab:branching-Ba+}
\begin{ruledtabular}
    \begin{tabular}{lllll}
    & \multicolumn{4}{c}{\ce{Ba+} upper state}\\
    \cline{2-5}
    & \multicolumn{1}{c}{$5d_{5/2}$} & \multicolumn{1}{c}{$6p_{1/2}$} & \multicolumn{1}{c}{$6p_{3/2}$} & \multicolumn{1}{c}{$7s$} \\
    \hline
    \smallspace
    $6s$  & 0.8302(9) & 0.7321(13) & 0.7421(13) & \multicolumn{1}{c}{$\sim10^{-14}$} \\
    & &{\sl0.731823(57)}&{\sl0.741716(71)}&\\
    \smallspace
    $5d_{3/2}$   & 0.1698(9) & 0.2679(13) & 0.02799(15) & $\sim5\times10^{-7}$ \\
    & &{\sl 0.268177(57)}&{\sl 0.028031(23)}&\\
    \smallspace
    $5d_{5/2}$   &  &  & 0.2299(11) & $\sim7\times10^{-7}$ \\
    & & &{\sl0.230253(61)}&\\
    \smallspace
    $6p_{1/2}$   &  &  & $\sim3\times10^{-10}$ & 0.34354(12) \\
    $6p_{3/2}$   &  &  &  & 0.65646(12)
    \end{tabular}
\end{ruledtabular}
\end{table}

\subsection{Lifetimes and branching fractions}

Here we consider the excited state lifetimes, and the branching fractions for their decays.
We determine the required transition rates using the standard formula (e.g., Ref.~\cite{Sobelman1992}), expressed in atomic units:\footnote{Note the M1 matrix elements are presented in the tables in units of the Bohr magneton; $\mu_B=\alpha/2$ in atomic units.}
\begin{equation}
    \gamma^{(k)}_{i\to f} = 
        \frac{2 (2k + 1)(k + 1)}{\left[(2k + 1)!!\right]^2\, k} \, (\omega \alpha)^{2k + 1} \frac{|\rme{f}{T^k}{i}|^2}{2J_i + 1},
\end{equation}
where $\omega$ is the transition frequency, and $k$ is the multipolarity ($k=1$ for E1 and M1, $k=2$ for E2).
To convert from atomic units (au), note that
\[
    1\,{\rm au} = \frac{\hbar}{E_H} = \frac{a_0}{c\alpha} = 2.41888...\times10^{-17}\,\si{s},
\]
where $E_H=27.211...\,$eV is the Hartree energy, $a_0$ is the Bohr radius, and $\alpha\approx1/137$ is the fine structure constant.
We use the matrix elements from~Table~\ref{tab:matel-summary}, and the experimental frequencies~\cite{NIST-ASD:2024, Kofford:2025, Holliman2020}.

For the uncertainties in both lifetimes and branching fractions, we account for the fact that the errors in the contributing matrix elements are highly correlated.
Therefore, we do not propagate through the uncertainties from the tables, but rather determine the uncertainties directly from the spread of values calculated at different levels, using the approach outlined in the methods section.

\begin{table}
\caption{Branching fractions for decays from the lowest few states of \ce{Ra+}. The $6d_{3/2}$ state decays only to $7s$. Experimental fractions~\cite{Fan:2019eze,Fan:2019} shown in italics}
\label{tab:branching-Ra+}
\begin{ruledtabular}
    \begin{tabular}{lllll}
        & \multicolumn{4}{c}{\ce{Ra+} upper state}\\
    \cline{2-5}
     & \multicolumn{1}{c}{$6d_{5/2}$} & \multicolumn{1}{c}{$7p_{1/2}$} & \multicolumn{1}{c}{$7p_{3/2}$} & \multicolumn{1}{c}{$8s$} \\
    \hline
    \smallspace
    $7s$ & 0.98488(9) & 0.91150(40) & 0.87740(62) & $\sim10^{-11}$ \\
    & &{\it 0.9104(7)}&{\it 0.87678(20)}&\\
        \smallspace
    $6d_{3/2}$ & 0.01512(9) & 0.08850(40) & 0.015329(45) & $\sim5\times10^{-7}$ \\
    & &{\it 0.0896(7)}&{\it 0.01563(21)}&\\
    \smallspace
    $6d_{5/2}$ &  &  & 0.10727(59) & $\sim8\times10^{-7}$ \\
    & & &{\it 0.10759(10)}&\\
    \smallspace
    $7p_{1/2}$ &  &  & $\sim5\times10^{-9}$ & 0.38310(70) \\
    $7p_{3/2}$ &  &  &  & 0.61690(70)
    \end{tabular}
\end{ruledtabular}
\end{table}

We present our calculated branching fractions for decays from the lowest few states of \ce{Ba+} and \ce{Ra+} in Tables~\ref{tab:branching-Ba+} and \ref{tab:branching-Ra+}, respectively.
The calculated lifetimes, with a comparison to experiment where available and to other theory values are presented in Tables~\ref{tab:lifetimes-Ba} and \ref{tab:lifetimes-Ra} for \ce{Ba+} and \ce{Ra+}, respectively.

The excellent agreement with experiment for the branching fractions in \ce{Ba+} indicates our uncertainties are likely overly conservative; the deviations from experiment are always less than three times the estimated theory uncertainty.
This is similar to what we found previously for the \ce{Sr+} branching fractions in Ref.~\cite{Roberts:2022lda}, where the deviations from experiment~\cite{Zhang2016} were an order of magnitude smaller than the theoretical uncertainties.
Theoretical values for the \ce{Ba+} $p_{3/2}$ branching fraction were also reported in Ref.~\cite{Zhang2020a}.
These are in agreement with our values, though our values have smaller uncertainty, and lie closer the experimental midpoints.

Experimental and theoretical branching fractions for the $p_{3/2}$ states of \ce{Ra+} have also been reported in Ref.~\cite{Fan:2019}.
In this case, there is only reasonable agreement with experiment, with a slight ($\sim$\,$1.4\sigma$) tension between our theory and the measured branching fraction for the $7p_{3/2}-6d_{3/2}$ decay channel.
At the same time, our theory results agree very well with other theory results, which use a very different coupled-cluster approach for the calculation~\cite{Fan:2019,Pal2009} (though are in significant disagreement with theory results from Ref.~\cite{Sahoo:2009-79.052512}).

Further, our calculated lifetime for the $7p_{3/2}$ state in \ce{Ra+} is in exceptional agreement with the experiment (Table~\ref{tab:lifetimes-Ra}), in significantly better agreement than previous calculations~\cite{Fan2022,Sahoo:2009-79.052512}.

\begin{table}
\caption{Calculated lifetimes for the lowest few states of \ce{Ba+}, and comparison with experiment and other theory.}
\label{tab:lifetimes-Ba}
\begin{ruledtabular}
    \begin{tabular}{llll}
    \ce{Ba+} & Final & Experiment & Other Theory \\
    \hline\\[-0.25cm]
    \multicolumn{4}{c}{{\em Long-lived states (s)}}\\
    \smallspace
    $5d_{3/2}$ & 82.4(10) & 79.8(4.6)~\cite{Yu1997} & 81.5(12)~\cite{Iskrenova-Tchoukova2008} \\
     &  & 89(16)~\cite{Gurell2007} & 80.09(71)~\cite{Sahoo2006} \\
     &  &  & 81.5~\cite{Dzuba:2001zz} \\
     \smallspace
    $5d_{5/2}$ & 30.63(37) & 30.14(40)~\cite{Zhang2020a} & 30.3(4)~\cite{Iskrenova-Tchoukova2008} \\
     &  & 31.2(9)~\cite{Auchter2014} & 29.86(30)~\cite{Sahoo2006} \\
     &  &  & 30.3~\cite{Dzuba:2001zz} \\
    \hline\\[-0.25cm]
    \multicolumn{4}{c}{{\em Short-lived states (ns)}}\\
    \smallspace
    $6p_{1/2}$ & 7.875(29) & 7.855(10)~\cite{Arnold2019} & 7.711~\cite{Arnold2019} \\
     &  & 7.9(1)~\cite{Pinnington1995} & 7.92(10)~\cite{Sahoo:2009-79.052512} \\
     &  & 7.92(8)~\cite{Kuske:1978-64.377} &  7.89~\cite{Dzuba:2001zz} \\
     \smallspace
    $6p_{3/2}$ & 6.300(23) & 6.271(8)\footnotemark[1]~\cite{Arnold2019} & 6.2615(72)~\cite{Zhang2020a} \\
     &  & 6.32(10)~\cite{Pinnington1995} & 6.30(17)~\cite{Sahoo:2009-79.052512} \\
     &  & 6.312(16)~\cite{Andra:1976} & 6.3~\cite{Dzuba:2001zz} \\
     &  & 6.31(7)~\cite{Kuske:1978-64.377} &  \\
     &  & 6.31(5)~\cite{Winter:1977} &  \\
    $7s$ & 5.121(68) &  &  \\
    $6d_{3/2}$ & 4.094(23) &  &  \\
    $6d_{5/2}$ & 4.339(26) &  &  \\
    $7p_{1/2}$ & 35.95(62) & 31.8(1.3)~\cite{Pinnington1995} &  \\
    $7p_{3/2}$ & 27.74(39) & 24.5(8)~\cite{Pinnington1995} &  \\
    $8s$ & 7.993(84) &  & 
    \end{tabular}
\end{ruledtabular}
\footnotemark[1]{Combination of experiment and theory}
\end{table}

\subsection{Extraction of E2 amplitudes from experiment}

Consider the ratio of reduced E2 matrix elements between the fine-structure partners of the lowest $s-d$ transitions, which is strongly dominated by its non-relativistic exact value:
\begin{equation}\label{eq:E2-ratio}
    R_{E2}(n) = 
    \left|\frac{\langle{(n+1)s_{1/2}}\|{E2}\|{nd_{5/2}}\rangle}
    {\langle{(n+1)s_{1/2}}\|{E2}\|{nd_{3/2}}\rangle}\right|
    \approx\sqrt{3/2},
\end{equation}
where $n=5$ for \ce{Ba+}, and $n=6$ for \ce{Ra+}.
Small deviations are due to relativistic effects, with only a very small corrections from many-body correlations.
Similarly, the M1 amplitude between the $d$-state fine-structure pair is dominated by its non-relativistic contribution,
\begin{equation}
    \label{eq:m1-nonrel}
    \langle{nd_{5/2}}\|{M1}\|{nd_{3/2}}\rangle \approx
    2\,\sqrt{\frac{3}{5}}\,\mu_B,
\end{equation}
with only very small contributions from many-body effects.

As such, these E2 ratios and M1 amplitudes can be determined theoretically with very high accuracy, as shown in Table~\ref{tab:Ratios}.
Therefore, these may be combined with experimental measurements to extract highly accurate values for the E2 amplitudes between the ground $s$ and lowest $d$ states for \ce{Ba+} and \ce{Ra+} from experiment.
There are two methods to do this.

\begin{table}
\caption{Calculated lifetimes for the lowest few states of \ce{Ra+}, and comparison with experiment and other theory.}
\label{tab:lifetimes-Ra}
\begin{ruledtabular}
    \begin{tabular}{lD{.}{.}{2.8}ll}
    \ce{Ra+} & \multicolumn{1}{c}{Final} & \multicolumn{1}{c}{Experiment} & \multicolumn{1}{c}{Other Theory} \\
    \hline\\[-0.25cm]
    \multicolumn{4}{c}{{\em Long-lived states (s)}}\\
    \smallspace
    $6d_{3/2}$ & 0.6452(41) & 0.642(9)~\cite{Li-arXiv:2501.18866} & 0.650(7)~\cite{Li-arXiv:2501.18866} \\
     &  &  & 0.642~\cite{Roberts:2013fxq} \\
     &  &  & 0.638(10)~\cite{Pal2009} \\
     &  &  & 0.627(4)~\cite{Sahoo2007} \\
     \smallspace
    $6d_{5/2}$ & 0.3062(33) & 0.3038(15)~\cite{Li-arXiv:2501.18866} & 0.307(3)~\cite{Li-arXiv:2501.18866} \\
     &  & 0.232(4)~\cite{Versolato:2010man} & 0.297(4)~\cite{Sahoo:2009-79.052512} \\
     &  &  & 0.302~\cite{Dzuba:2001zz} \\
    \hline\\[-0.25cm]
    \multicolumn{4}{c}{{\em Short-lived states (ns)}}\\
    \smallspace
    $7p_{1/2}$ & 8.829(23) &  & 8.57(10)~\cite{Sahoo:2009-79.052512} \\
    &  &  & 8.72~\cite{Pal2009} \\
    $7p_{3/2}$ & 4.767(9) & 4.78(3)~\cite{Fan2022} & 4.67(9)~\cite{Sahoo:2009-79.052512} \\
    &  &  & 4.73~\cite{Pal2009} \\
    $8s$ & 5.570(42) &  &  \\
    $7d_{3/2}$ & 4.351(22) &  &  \\
    $7d_{5/2}$ & 4.978(21) &  &  \\
    $8p_{1/2}$ & 51.28(47) &  &  \\
    $8p_{3/2}$ & 20.07(28) &  &  \\
    $9s$ & 9.256(31) &  & 
    \end{tabular}
\end{ruledtabular}
\end{table}

\begin{table}
\caption{Calculations of the $s-d$ E2 ratio~\eqref{eq:E2-ratio}, and the $d_{3/2}-d_{5/2}$ M1 reduced matrix element~\eqref{eq:m1-nonrel} for \ce{Ba+} and \ce{Ra+}. These may be combined with lifetime measurements of the $d$ states to extract the E2 matrix elements.}
\label{tab:Ratios}
\begin{ruledtabular}
    \begin{tabular}{lD{.}{.}{0.6}D{.}{.}{0.6}D{.}{.}{0.5}D{.}{.}{0.7}}
     & \multicolumn{2}{c}{\ce{Ba+}} & \multicolumn{2}{c}{\ce{Ra+}} \\
     \cline{2-3}\cline{4-5}
     & \multicolumn{1}{l}{$R_{E2}$} & \multicolumn{1}{l}{$M1$ ($\mu_B$)} & \multicolumn{1}{l}{$R_{E2}$} & \multicolumn{1}{l}{$M1$ ($\mu_B$)} \\
    \hline
    \smallspace
    Non-rel & 1.22474 & 1.54919 & 1.22474 & 1.54919 \\
    HF & 1.24525 & 1.54889 & 1.26114 & 1.54780 \\
    RPA & 1.24782 & 1.54976 & 1.26808 & 1.55497 \\
    $\Sigma^{(2)}$ & 1.25521 & 1.54971 & 1.29046 & 1.55465 \\
    $\Sigma^{(\infty)}$ & 1.25381 & 1.54973 & 1.28525 & 1.55490 \\
    \smallspace
    $\lambda_{\rm scale}$ & -0.00024 & 0 & -0.00048 & -0.00003 \\
    Breit & -0.00070 & -0.00004 & -0.00071 & -0.00007 \\
    QED & 0.00012 & 0 & 0.00034 & 0 \\
    SR+N & -0.00155 & -0.00041 & -0.00511 & -0.00369 \\
    \smallspace
    Final & 1.25144(60) & 1.54928(21) & 1.2793(16) & 1.5511(18)
    \end{tabular}
\end{ruledtabular}
\end{table}

\begin{figure}
    \includegraphics[width=0.4\textwidth]{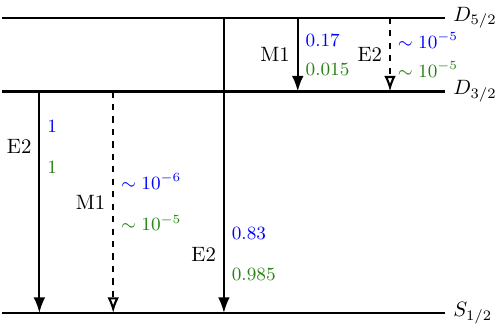}
    \caption{Level scheme for the long-live states, showing approximate branching ratios for \ce{Ba+} (blue) and \ce{Ra+} (green).}
    \label{fig:Levels}
\end{figure}

\paragraph*{\bf Method I---}
In the first method, we exploit the fact that, for both ions, the $d_{3/2}$ state decays exclusively to the ground $s$ state, with the E2 channel dominating over M1 by about five orders of magnitude (see Fig.~\ref{fig:Levels}).
Thus, the $d_{3/2}$ lifetime measurement allows a clean extraction of the $s-d_{3/2}$ E2 matrix element.
From the E2 ratio~\eqref{eq:E2-ratio}, we can then infer the value of the $s-d_{5/2}$ matrix element.
At the same time, the $d_{5/2}$ state decays to $s_{1/2}$ (via a now-known E2 matrix element), and to $d_{3/2}$ via M1 (with a negligible E2 contribution, Fig.~\ref{fig:Levels}).
Therefore, the $d_{5/2}$ lifetime measurement further allows extraction of the $d_{3/2}-d_{5/2}$ M1 matrix element.
Note that the uncertainty in Method 1 is limited by the $d_{3/2}$ lifetime measurements, whereas the $d_{5/2}$ lifetime is known significantly more precisely for both considered ions.

\paragraph*{\bf Method II---}
In Method 2, we take advantage of the fact that the $d_{3/2}-d_{5/2}$ M1 matrix element~\eqref{eq:m1-nonrel} can be calculated with high accuracy.
Further, this decay is mainly due to the E2 channel, with the M1 channel contributing at the $\lesssim20\%$ level (see Table~\ref{tab:branching-Ra+}).
Thus, we can reverse the above procedure: use the measured $d_{5/2}$ lifetime along with the calculated M1 value to determine the $s-d_{5/2}$ matrix element with high precision, and then, using the known ratio, infer the $s-d_{3/2}$ matrix element.

\begin{figure*}
    \centering
    \includegraphics[width=0.485\linewidth]{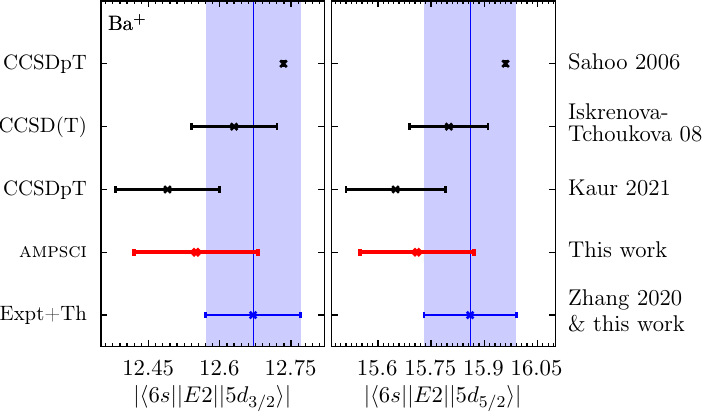}~
    \includegraphics[width=0.485\linewidth]{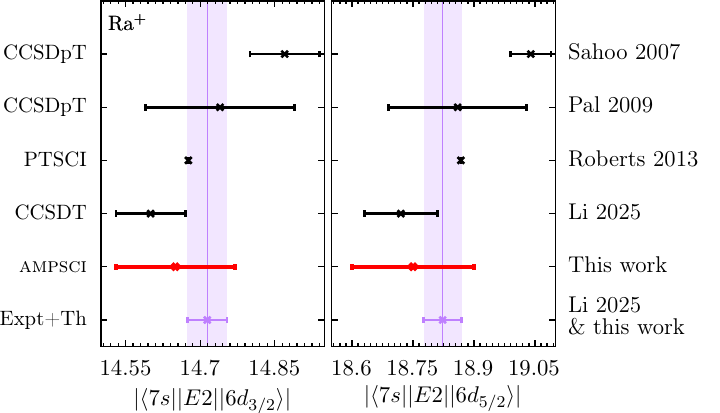}
    \caption{Comparison of reduced electric quadrupole (E2) transition amplitudes for \ce{Ba+} (left) and \ce{Ra+} (right). The shaded region shows the value extracted  in this work from experiment (Zhang {\sl et al.}~\cite{Zhang2020a} and Li~{\sl et al.}~\cite{Li-arXiv:2501.18866}), the red points are theory values from this work, and the black points are other high-precision theory values (Sahoo {\sl et al.}~\cite{Sahoo2006,Sahoo2007},  Iskrenova-Tchoukova {\sl et al.}~\cite{Iskrenova-Tchoukova2008}, Kaur {\sl et al.}~\cite{Kaur:2020nsc}, and Roberts {\sl et al.}~\cite{Roberts:2013fxq}).
    }
    \label{fig:E2}
\end{figure*}

For the present data, method II allows much higher precision extraction of the E2 matrix elements.
These are presented in Table~\ref{tab:E2-extracted}.
In all cases, the precision is high, and the uncertainty is dominated by experiment.
The uncertainty for the E2 matrix elements of \ce{Ra+} reaches the 0.2\% level, while for \ce{Ba+} it is closer to 1\%.
The agreement between these values and our theory calculations is excellent, as shown in Fig.~\ref{fig:E2}.
We also show the extracted M1 matrix elements (using Method I); despite the very large experimental uncertainties, these are in agreement with theory and serve as a consistency check.

\begin{table}
\caption{Values for the E2 ($ea_0^2$) and M1 ($\mu_B$) reduced matrix elements extracted from experiment for \ce{Ba+} ($n=5$) and \ce{Ra+} ($n=6$). These are determined from the lifetime measurements in Tables~\ref{tab:lifetimes-Ba} and \ref{tab:lifetimes-Ra}, and the theory values from Table~\ref{tab:Ratios}.
The resulting uncertainties are dominated by the lifetime measurements; the theory contribution to the uncertainty is also shown when it is significant.}
\label{tab:E2-extracted}
\begin{ruledtabular}
    \begin{tabular}{c D{.}{.}{2.9} D{.}{.}{2.13}}
    &\multicolumn{1}{c}{\ce{Ba+}}&\multicolumn{1}{c}{\ce{Ra+}}\\
    \hline
    \smallspace
    $|\langle{(n+1)s}\|E2\|{nd_{3/2}}\rangle|$ & 12.67(10)_e & 14.714(36)_e(18)_t\\
    $|\langle{(n+1)s}\|E2\|{nd_{5/2}}\rangle|$ & 15.86(13)_e & 18.823(47)_e\\
    $|\langle{nd_{3/2}}\|M1\|{nd_{5/2}}\rangle|$ & 1.7(6)_e(1)_t & 1.53(25)_e\\
    \end{tabular}
\end{ruledtabular}
\end{table}

\section{Hyperfine Constants}

The hyperfine $A$ constants for all considered states, including a breakdown of contributions at different approximations, are presented in the appendix (Table~\ref{tab:hfs_BaRa+}).
All values are in agreement with available experiment, indicating the theoretical uncertainties are reasonable.

\begin{table}[]
\caption{Summary of calculated hyperfine $A$ constants (MHz) for the ground states of \ce{Ba+} and \ce{Ra+} in the pointlike nuclear magnetization model, and comparison with experiment; nuclear $g$ factors from Ref.~\cite{Mertzimekis:2016,*IAEAonline} are also shown. 
From this, we extract accurate determinations of the relative Bohr-Weisskopf effects, $\epsilon$ [Eq.~\eqref{eq:BW}], for the $s$, $p$ and $d$ states; the uncertainties are dominated by atomic theory.}
\label{tab:hfs_summary}
\begin{ruledtabular}
    \begin{tabular}{lllll}
     & \multicolumn{1}{c}{\ce{^135Ba+}} & \multicolumn{1}{c}{\ce{^137Ba+}} & \multicolumn{1}{c}{\ce{^223Ra+}} & \multicolumn{1}{c}{\ce{^225Ra+}} \\
     \hline\smallspace
    $A_{\rm Th.}^{\rm point}$ & 3627(13) & 4057(15) & 3544(29) & $-$28830(240) \\
    $A_{\rm Expt.}$ & 3591.670...\footnotemark[1] & 4018.870...\footnotemark[2] & 3404.0(19)\footnotemark[3] & $-$27684.511...\footnotemark[4] \\
    \smallspace
    $g$~\cite{Mertzimekis:2016,*IAEAonline} & 0.5587(1) & 0.6250(1) & 0.1795(5) & $-$1.460(4) \\
    \hline~\\
    \multicolumn{5}{c}{{\it Extracted relative BW effects (\%)}}\\
    \smallspace
    $\epsilon(s)$ & $-$0.96(36) & $-$0.94(36) & $-$3.94(82) & $-$3.96(81) \\
    $\epsilon(p_{1/2})$ & $-$0.06(3) & $-$0.06(3) & $-$1.25(32) & $-$1.26(32) \\
    $\epsilon(p_{3/2})$ & $-$0.25(11) & $-$0.24(11) & $-$1.05(27) & $-$1.06(27) \\
    $\epsilon(d_{3/2})$ & $-$0.6(3) & $-$0.6(3) & $-$3.7(10) & $-$3.7(10) \\
    $\epsilon(d_{5/2})$ & $-$1.7(8) & $-$1.6(8) & $-$5.2(14) & $-$5.2(13)
    \end{tabular}
\end{ruledtabular}
Experimental references:~\footnotemark[1]{\cite{Trapp:2000}}
\footnotemark[2]{\cite{Blatt:1982}}
\footnotemark[3]{\cite{Neu:1988wt}}
\footnotemark[4]{\cite{Ready:2024xxx}}.
\end{table}

For the hyperfine constants, there is an important contribution from the finite nuclear magnetization distribution, known as the Bohr-Weisskopf (BW) effect.
This depends on nuclear structure, and is significant at the current level of theoretical accuracy~\cite{Ginges:2017rmj}; nuclear uncertainties entering via the BW effect can cloud comparisons between atomic theory and experiment.
For the values in Table~\ref{tab:hfs_BaRa+}, we use a simple single particle model (see, e.g., Ref.~\cite{Volotka:2008.78.062507}) to estimate the BW effect independently from experiment, with an assumed 50\% uncertainty.
Note that this model has been tested for other atoms, and shown to be reasonably accurate~\cite{Roberts:2021tla,Sanamyan:2022wgr}.

It is convenient to express the hyperfine constant as
\begin{equation}\label{eq:BW}
    A = A_0\,(1+\epsilon) + \delta A_{\rm QED},
\end{equation}
where $A_0$ includes the effects of the finite nuclear charge distribution, but assumes a pointlike nuclear magnetization distribution, and $\delta A_{\rm QED}$ is the QED correction (which we take from Ref.~\cite{Ginges:2017fle}).
The relative BW corrections, $\epsilon$, depend on the electronic angular quantum numbers, though, for $s$- and $p$- states, are independent of the principle quantum number~\cite{Grunefeld:2019.100.042506}.
As can be seen from this equation, by comparing the calculations with experiment, a value of $\epsilon$ can be extracted from the experiment so long as the atomic uncertainty is small enough.
This is performed for the ground states of isotopes of \ce{Ba+} and \ce{Ra+} in Table~\ref{tab:hfs_summary}.
The uncertainty in the extracted factors is dominated by the atomic theory.
The similar process was performed for the $7s$ ground state of \ce{^225Ra+} in Ref.~\cite{Skripnikov:2020zst} using an older experimental value.

Further, it turns out that the ratios of Bohr-Weisskopf effects for different states of the same atom are essentially independent of the nuclear magnetization model, and are also very insensitive to electron correlation effects~\cite{Roberts:2021fpk}.
Therefore, we may use ratios calculated using a simple single-particle nuclear model to determine the BW effect for higher angular states from the value extracted for the ground state; these are also presented in Table~\ref{tab:hfs_summary}.
Notice that these corrections are large for both ions, and particularly so for \ce{Ra+}, as previously noted~\cite{Skripnikov:2020zst}.

The extracted Bohr-Weisskopf effect for \ce{^225Ra+} (from the measurement of the $7s$ hyperfine splitting) improves the accuracy of the theory value for the $7p_{1/2}$ state, further indicating that the extracted BW effects are accurate.
For example, in the pointlike approximation, we calculate $A_{7p_{1/2}} = -5554(83)$\,MHz (including the QED correction).
The BW effect extracted in Table~\ref{tab:hfs_summary} implies a correction of $+70(18)$\,MHz, bringing the predicted value to $-5484(85)$\,MHz, in excellent agreement with experiment $-5447(4)$~\cite{Ready:2024xxx}.
The corrections to the $d$ states also lead to improved agreement, though the theory uncertainty is much larger for those states.

\section{Polarizabilities}

Tables \ref{tab:Alpha0} and \ref{tab:Alpha2} present our calculations for the scalar and tensor polarizabilities, respectively, along with comparison with experimental data where available and calculations of other groups. 
As can be seen in these two tables, our calculations are in reasonable agreement with previous calculations.
Structure radiation corrections are very small for the ground states, though become significant (few \%) for the excited states, which is reflected in the increased uncertainty.
For the tensor polarizabilities (Table \ref{tab:Alpha2}), theoretical values deviate substantially across different groups, likely due to very significant role of correlation corrections.

\begin{table}
\caption{Calculated static scalar polarizabilities (${a_0^3}$) for the lowest \ce{Ba+} and \ce{Ra+} valence states.}
\label{tab:Alpha0}
\begin{ruledtabular}
\begin{tabular}{lD{.}{.}{5.2}D{.}{.}{5.6}lr}
    State & \multicolumn{1}{c}{RHF} & \multicolumn{1}{c}{Final} & \multicolumn{1}{c}{Other} & Expt. \cite{Snow2007} \\ 
    \hline
    \multicolumn{5}{l}{\ce{Ba+}} \\
    \smallspace
    6$s$ & 185.50 & 123.26(45) & 124.3(10)~\cite{Sahoo:2009-80.062506} & 123.88(5) \\
    & & &124.15~\cite{Iskrenova-Tchoukova2008}&\\
    \smallspace
    6$p_{1/2}$ & -30.1 & 19.5(25) &   &  \\
    6$p_{3/2}$ & 4.3 & 44.0(25) &  &  \\
    5$d_{3/2}$ & 91.3 & 49.5(17) &  &  \\
    5$d_{5/2}$ & 88.8 & 49.6(17) & &  \\
    \hline
    \multicolumn{5}{l}{\ce{Ra+}} \\
    \smallspace
    7$s_{1/2}$ & 163.81 & 105.02(61) & 104.5(15)~\cite{Sahoo:2009-80.062506} & \\
    & & & 103.21~\cite{Wu:2016-25.093101} &  \\
    & & & 106.5~\cite{Safronova2007} &  \\
    \smallspace
    7$p_{1/2}$ & -210.8 & -43.7(39) &  & \\
    7$p_{3/2}$ & -24.1 & 50.2(32) &  &  \\
    \smallspace
    6$d_{3/2}$ & 200.0 & 89.7(28) & 83.71(77)~\cite{Sahoo:2009-80.062506} & \\
    & & &88.6~\cite{Wu:2016-25.093101}  &  \\
    \smallspace
    6$d_{5/2}$ & 153.8 & 82.6(23) & 82.38(70)~\cite{Sahoo:2009-80.062506} \\
    & & & 82.2~\cite{Wu:2016-25.093101} & 
\end{tabular}
\end{ruledtabular}
\end{table}

\begin{table}
\caption{
Calculated static tensor polarizabilities (${a_0^3}$) for the lowest \ce{Ba+} and \ce{Ra+} states.}
\label{tab:Alpha2}
\begin{ruledtabular}
\begin{tabular}{lD{.}{.}{3.2} D{.}{.}{3.5}l}
    State & \multicolumn{1}{c}{RHF} &  \multicolumn{1}{c}{Final} & \multicolumn{1}{c}{Other} \\
    \hline
    \multicolumn{4}{l}{\ce{Ba+}} \\
    \smallspace
    6$p_{3/2}$  & 23.44 &  5.18(35)  &   \\
    5$d_{3/2}$  & -45.5 &  -21.5(11)  &  \\
    5$d_{5/2}$  & -58.63 &  -28.7(14)  &  $-$29.43(52)~\cite{Jiang2021} \\
    \hline
    \multicolumn{4}{l}{\ce{Ra+}}   \\
    \smallspace
    7$p_{3/2}$  & -1.97 &  -17.79(52)  &   \\
    6$d_{3/2}$  & -128.7 &  -48.7(20)  &  $-$50.23(43)~\cite{Sahoo:2009-80.062506} \\
     &  &  &  $-$49.1~\cite{Wu:2016-25.093101} \\
     \smallspace
    6$d_{5/2}$  & -108 &  -51.1(19)  &  $-$52.60(45)~\cite{Sahoo:2009-80.062506} \\
     &  &  &  $-$53.0~\cite{Wu:2016-25.093101}\\
\end{tabular}
\end{ruledtabular}
\end{table}


\begin{figure*}
    \centering
    \includegraphics[width=0.48\textwidth]{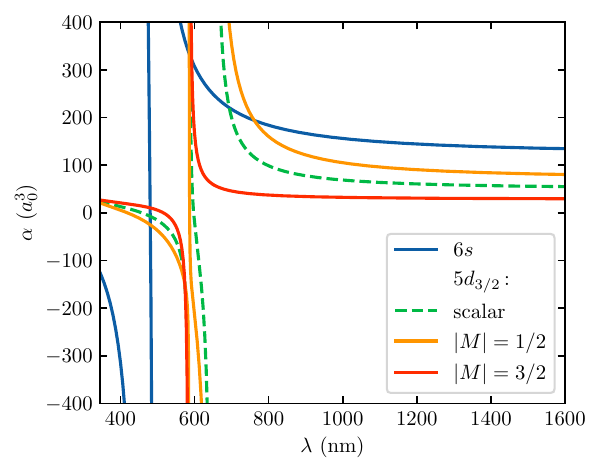}
    \includegraphics[width=0.48\textwidth]{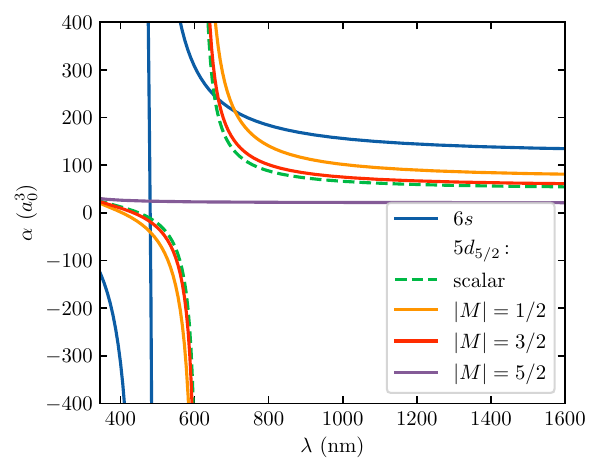}
    \caption{Dynamic polarizability of the $6s-5d_{3/2}$ (left) and $6s-5d_{5/2}$ (right) transitions in \ce{Ba+}.}
    \label{fig:Ba2_dp}
\end{figure*}

\begin{figure*}
    \centering
    \includegraphics[width=0.48\textwidth]{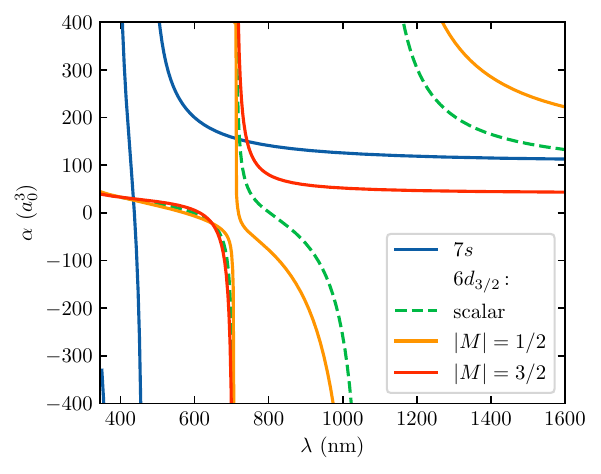}
    \includegraphics[width=0.48\textwidth]{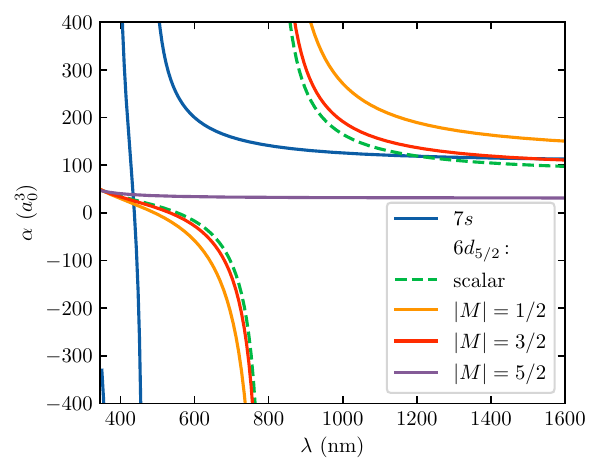}
    \caption{Dynamic polarisability for the $7s-6d_{3/2}$ (left) and $7s-6d_{5/2}$ (right) transitions in \ce{Ra+}.}
    \label{fig:7s-6dp}
\end{figure*}

For linearly polarized light of angular frequency $\omega$, the dynamic polarizability of the state $v$ may be expressed in terms of scalar and tensor components~\cite{Mitroy2010}:
\begin{equation} \label{eq:pol}
   \alpha (\omega) = \alpha_0(\omega) + \alpha_2(\omega)\,\frac{3M_v^2-J_v(J_v+1)}{J_v\,(2J_v - 1)} \eta \, ,
\end{equation}
where $J_v$ and $M_v$ are the total angular momentum and its projection along the $z$ quantization axis, 
and $\eta = \tfrac{1}{2}(3\cos^2\theta-1)$ with $\theta$ the angle between the light polarization and the $z$-axis.
The scalar polarizability is
\begin{equation}
    \alpha_0(\omega) = \frac{2}{3\,(2J_v+1)}\,\sum_{n}\frac{\Delta\varepsilon_{nv}\,|\langle v||d||n\rangle|^2}{\Delta\varepsilon^{2}_{nv}- \omega^2}\, ,
\end{equation}
where $\Delta\varepsilon_{nv}=\varepsilon_n - \varepsilon_v$ are the transition energies.
Similarly, the tensor polarizability is
\begin{equation}
    \alpha_2(\omega) = C\,\sum_{n}(-1)^{J_v+J_n} \begin{Bmatrix} J_v & 1 & J_n \\ 1 & J_n & 2\end{Bmatrix}
    \frac{\Delta\varepsilon_{nv}\,|\langle v||d||n\rangle|^2}{\Delta\varepsilon^{2}_{nv}-\omega^2}\,,\\
\end{equation}
where
{
\begin{equation}
    C=4\left(\frac{5J_v(2J_v-1)}{6(J_v+1)(2J_v+1)(2J_v+3)}\right)^{1/2} \, ,\nonumber
\end{equation}}
and $\{:::\}$ is a 6$j$-symbol.
Of interest are the magic wavelengths~\cite{Katori:1999}, where the polarizabilities of the two states in the transition become equal, so that the light shift for the transition vanishes.
These are of interest experimentally, and also as a precise tool for testing atomic theory.

We calculate the dynamic polarizabilities and magic wavelengths for the relevant clock transitions. 
Our calculations were evaluated at the level of the scaled all-orders method ($\lambda\Sigma^{(\infty)}$).
For the $d$ states, we include the tensor contributions.
This is the same method used by one of us recently for the calculating the static and dynamic polarizabilities in Rb~\cite{HamiltonRb2022}.
Comparison to experiment in that work demonstrated excellent agreement with experiment for the differential polarizabilities, both near and far from the resonances, and for the magic wavelengths.
The dynamic polarizabilities for \ce{Ba+} and \ce{Ra+} are shown in Figs.~\ref{fig:Ba2_dp} and \ref{fig:7s-6dp}, respectively.

The relevant magic wavelengths are presented in Table~\ref{tab:MagicLbd_BaRa}.
We extract both the full magic wavelengths (i.e., including the tensor contribution for particular projection $M=J_z$), as well as the scalar-only values, which can be selected experimentally by averaging over projections~\cite{Chanu2020}.
The structure radiation contribution is the dominant source of theoretical uncertainty, particularly for the tensor polarizability contributions.
The scalar magic wavelength for the \ce{Ba+} $6s-5d_{5/2}$ transition around 650\,nm is excellent agreement with high-precision experiment~\cite{Chanu2020}, indicating excellent theoretical accuracy, and again suggesting our theoretical uncertainties are conservative.

The magic wavelengths for the $7s-6d_{3/2}$ and $7s-6d_{5/2}$ transitions in \ce{Ra+} that occur around $\sim$\,430\,nm occur close to an $s$-state resonance, and as such have particularly small contributions from the tensor polarizability.
This may have experimental advantages due to suppressed sensitivity to experimental parameters such as the angle of the polarization.

The magic wavelengths for the $7s-6d_{3/2}$ and $7s-6d_{5/2}$ transitions in \ce{Ra+} that occur around $\sim$\,430\,nm lie close to an $s$-state resonance, and therefore have particularly small contributions from the tensor polarizability (which is zero for $s$-states).
This may offer experimental advantages due to reduced sensitivity to parameters such as the polarization angle, $\theta$.
At the same time, being close to the resonance means the polarizability will have large sensitivity to variations in the frequency.
In contrast, the magic wavelength for \ce{Ra+} $7s-6d_{5/2}$ transition that occurs above $\sim$\,1600\,nm is very shallow (i.e., occurs far from a resonance), and is therefore very sensitive to the details of the calculation, and so an accurate value cannot be easily determined.
At the same time, it would be very intensive to variations in the frequency.

\begin{table}
\caption{Relevant magic wavelengths (nm) for the \ce{Ba+} and \ce{Ra+} transitions.}
\label{tab:MagicLbd_BaRa}
\begin{ruledtabular}
\begin{tabular}{lcD{,}{}{4.6}l}
    & &\multicolumn{2}{c}{$\lambda_{m}$(nm)}\\
    \cline{3-4}
    Transition & $|M|$ & \multicolumn{1}{c}{This work} & \multicolumn{1}{c}{Other} \\
    \hline
    \multicolumn{1}{l}{\ce{Ba+}}& & &\\
    \smallspace
    $6s-5d_{3/2}$ & Scalar\footnotemark[1] & 692,.7(8) &   \\
    $6s-5d_{3/2}$ & 1/2 & 757,.2(26) &   \\ 
    \smallspace
    $6s-5d_{5/2}$ & Scalar & 652,.3(10) & 652.9130(40)\footnotemark[2]~\cite{Chanu2020} \\
    $6s-5d_{5/2}$ & 1/2 & 705,(3) & 715.6(138)~\cite{ShanShan_2024}  \\  
     &  &  & 707.9(33)~\cite{Jiang2021} \\
    $6s-5d_{5/2}$ & 3/2 & 662,.8(13) & 666.7(66)~\cite{ShanShan_2024} \\
     &  &  & 663.6(14)~\cite{Jiang2021} \\
    \hline
    \multicolumn{4}{l}{\ce{Ra+}}\\
    \smallspace
    $7s - 6d_{3/2}$  & Scalar & 434,.5(10) & \\ 
    $7s - 6d_{3/2}$  & 1/2 & 435,.4(5) &  \\ 
    $7s - 6d_{3/2}$  & 3/2 & 433,.5(20) &  \\ 
    \smallspace
    $7s - 6d_{3/2}$  & Scalar & 722,.4(10) & \\ 
    $7s - 6d_{3/2}$  & 1/2 & 710,.335(50) &  \\ 
    $7s - 6d_{3/2}$  & 3/2 & 749,(8) &  \\ 
    \smallspace
    $7s - 6d_{5/2}$  & Scalar & 433,.7(20) & \\ 
    $7s - 6d_{5/2}$  & 1/2 & 434,.5(20) &  \\ 
    $7s - 6d_{5/2}$  & 3/2 & 433,.8(20) &  \\ 
    $7s - 6d_{5/2}$  & 5/2 & 432,.8(20) &  \\ 
\end{tabular}
\end{ruledtabular}
\footnotemark[1]{Scalar-only value.}
\footnotemark[2]{Experimental}
\end{table}

\begin{table*}[t!]
\caption{Reduced matrix elements for electric dipole (E1) transitions of \ce{Ba+} and \ce{Ra+}.
The `Final' column includes the Breit, QED, scaling, and structure radiation and normalization corrections.
Transitions between the lower $6s,\,7s,\,6p,\,7p$, and $5d$ states of \ce{Ba+} (and similarly with $n+1$ for \ce{Ra+}) were recently presented in Ref.~\cite{Roberts:2022lda}.}
\label{tab:E1_BaRa}
\begin{ruledtabular}
\begin{tabular}{rlD{.}{.}{1.4}D{.}{.}{1.4}D{.}{.}{1.4}D{.}{.}{1.8}|rlD{.}{.}{1.4}D{.}{.}{1.4}D{.}{.}{1.4}D{.}{.}{1.8}}
     \multicolumn{2}{c}{\ce{Ba+}}  & \multicolumn{1}{c}{HF} & \multicolumn{1}{c}{RPA} & \multicolumn{1}{c}{$\Sigma^{(\infty)}$} & \multicolumn{1}{c}{Final} &
    \multicolumn{2}{c}{\ce{Ra+}}  & \multicolumn{1}{c}{HF} & \multicolumn{1}{c}{RPA} & \multicolumn{1}{c}{$\Sigma^{(\infty)}$} & \multicolumn{1}{c}{Final} \\
     \hline
     \smallspace
    8$p_{1/2}$ & 6$s$ & 0.0071 & 0.1050 & 0.1151 & 0.101(6) & 9$p_{1/2}$ & 7$s$ & 0.0239 & -0.0904 & -0.1022 & -0.092(8) \\
    8$p_{3/2}$ & 6$s$ & 0.0786 & -0.0619 & -0.0714 & -0.052(8) & 9$p_{3/2}$ & 7$s$ & 0.2814 & 0.1137 & 0.1106 & 0.123(11) \\
    8$p_{1/2}$ & 7$s$ & 0.1989 & 0.1249 & 0.1133 & 0.125(5) & 9$p_{1/2}$ & 8$s$ & 0.2753 & 0.1876 & 0.1645 & 0.159(13) \\
    8$p_{3/2}$ & 7$s$ & 0.5675 & 0.4658 & 0.4663 & 0.480(8) & 9$p_{3/2}$ & 8$s$ & 1.1405 & 1.0301 & 1.0413 & 1.029(15) \\
    6$p_{1/2}$ & 8$s$ & 0.7015 & 0.7251 & 0.7112 & 0.7014(31) & 7$p_{1/2}$ & 9$s$ & 0.7159 & 0.7327 & 0.7154 & 0.7067(28) \\
    7$p_{1/2}$ & 8$s$ & 5.0876 & 5.1198 & 4.9576 & 4.953(17) & 8$p_{1/2}$ & 9$s$ & 5.2274 & 5.2598 & 5.0383 & 5.030(16) \\
    8$p_{1/2}$ & 8$s$ & 11.917 & 11.855 & 11.571 & 11.546(13) & 9$p_{1/2}$ & 9$s$ & 11.8734 & 11.800 & 11.511 & 11.502(8) \\
    6$p_{3/2}$ & 8$s$ & 1.0294 & 1.0542 & 1.0351 & 1.0216(40) & 7$p_{3/2}$ & 9$s$ & 1.0784 & 1.0730 & 1.0475 & 1.0364(43) \\
    7$p_{3/2}$ & 8$s$ & 7.8099 & 7.8471 & 7.6563 & 7.645(26) & 8$p_{3/2}$ & 9$s$ & 9.2439 & 9.2585 & 9.0508 & 9.029(25) \\
    8$p_{3/2}$ & 8$s$ & 16.551 & 16.470 & 16.049 & 16.015(21) & 9$p_{3/2}$ & 9$s$ & 15.713 & 15.632 & 15.145 & 15.156(19) \\
    8$p_{1/2}$ & 5$d_{3/2}$ & 0.1956 & 0.1072 & 0.1258 & 0.129(11) & 9$p_{1/2}$ & 6$d_{3/2}$ & 0.0868 & -0.0340 & 0.0048 & 0.011(16) \\
    8$p_{3/2}$ & 5$d_{3/2}$ & 0.1019 & 0.0643 & 0.0727 & 0.0747(45) & 9$p_{3/2}$ & 6$d_{3/2}$ & 0.0942 & 0.0448 & 0.0633 & 0.067(7) \\
    6$p_{1/2}$ & 6$d_{3/2}$ & 5.1408 & 5.0035 & 4.8983 & 4.884(18) & 7$p_{1/2}$ & 7$d_{3/2}$ & 4.5268 & 4.4345 & 4.3518 & 4.344(26) \\
    7$p_{1/2}$ & 6$d_{3/2}$ & 9.1892 & 9.1025 & 8.6704 & 8.627(15) & 8$p_{1/2}$ & 7$d_{3/2}$ & 10.207 & 10.058 & 9.580 & 9.526(17) \\
    8$p_{1/2}$ & 6$d_{3/2}$ & 0.0878 & 0.0425 & 0.1181 & 0.122(7) & 9$p_{1/2}$ & 7$d_{3/2}$ & 0.5140 & 0.5898 & 0.4521 & 0.456(31) \\
    6$p_{3/2}$ & 6$d_{3/2}$ & 2.4458 & 2.3790 & 2.3391 & 2.330(8) & 7$p_{3/2}$ & 7$d_{3/2}$ & 2.4883 & 2.4249 & 2.4090 & 2.399(9) \\
    7$p_{3/2}$ & 6$d_{3/2}$ & 4.0198 & 3.9869 & 3.7851 & 3.767(6) & 8$p_{3/2}$ & 7$d_{3/2}$ & 4.3307 & 4.2842 & 4.0440 & 4.023(6) \\
    8$p_{3/2}$ & 6$d_{3/2}$ & 0.1297 & 0.1122 & 0.1473 & 0.1482(21) & 9$p_{3/2}$ & 7$d_{3/2}$ & 0.0936 & 0.0685 & 0.1333 & 0.129(11) \\
    8$p_{3/2}$ & 5$d_{5/2}$ & 0.2976 & 0.1907 & 0.2152 & 0.218(14) & 9$p_{3/2}$ & 6$d_{5/2}$ & 0.2630 & 0.1344 & 0.1836 & 0.184(21) \\
    6$p_{3/2}$ & 6$d_{5/2}$ & 7.2533 & 7.0591 & 6.9377 & 6.915(22) & 7$p_{3/2}$ & 7$d_{5/2}$ & 7.2489 & 7.0733 & 6.9966 & 6.978(25) \\
    7$p_{3/2}$ & 6$d_{5/2}$ & 12.2169 & 12.1157 & 11.5226 & 11.465(20) & 8$p_{3/2}$ & 7$d_{5/2}$ & 13.3725 & 13.2284 & 12.5749 & 12.508(24) \\
    8$p_{3/2}$ & 6$d_{5/2}$ & 0.3184 & 0.2654 & 0.3685 & 0.371(8) & 9$p_{3/2}$ & 7$d_{5/2}$ & 0.0897 & 0.0145 & 0.1865 & 0.168(39)
\end{tabular}
\end{ruledtabular}
\end{table*}

\section{Conclusion}

In this work, we have performed detailed high-accuracy calculations of electric dipole (E1), electric quadrupole (E2), and magnetic dipole (M1) matrix elements, excited-state lifetimes, polarizabilities, magic wavelengths, and hyperfine structure constants for \ce{Ba+} and \ce{Ra+}. 
The calculations were compared to precise experimental values, where available, demonstrating both high accuracy, and the robustness of the method used to estimate the theoretical uncertainties.
We extracted improved values for key E2 transition matrix.
We also extracted Bohr-Weisskopf corrections, which describe the finite distribution of the magnetic moment across the nuclei.
These are larger-than-typical for both \ce{Ba} and \ce{Ra}, and are crucial for accurate determinations of the hyperfine structure.
Together, these provide new insights into both atomic and nuclear structure, and offer important theoretical support for current and future programs in atomic parity violation, optical clock development, and searches for physics beyond the Standard Model.

\acknowledgements

This work was supported by Australian Research Council (ARC) DECRA Fellowship DE210101026, and Discovery Project No.~DP230101685.
Calculations were performed with the \textit{UQ Research Computing Centre}.

\onecolumngrid
\appendix

\section*{Appendix}

Table~\ref{tab:E1_BaRa} presented the reduced matrix elements for electric dipole (E1) transitions of \ce{Ba+} and \ce{Ra+}.
Tables~\ref{tab:M1E2-ba} and \ref{tab:M1E2-ra} present the reduced matrix elements for the magnetic dipole (M1) and electric quadrupole (E2) transitions between the lowest few states of \ce{Ba+} and \ce{Ra+}, respectively, showing the contributions from different levels of approximation.
Table~\ref{tab:hfs_BaRa+} presents the hyperfine A constants for \ce{^135Ba+} and \ce{^223Ra+}, including the breakdown of the calculations at different approximations, and comparison with experiment.
The ``subtotal'' column refers to the calculations performed in the pointlike nuclear magnetization model (finite nuclear charge effects are accounted for).
The theoretical values for the hyperfine constants of other isotopes can be found by rescaling the pointlike values by the respective nuclear $g$-factors (the small correction due to the change in nuclear charge radius, the relative Breit-Rosenthal effect, is very small).
Note that we use the simple single particle model (see, e.g., Refs.~\cite{Volotka:2008.78.062507, Roberts:2020vef}) to estimate the BW effect independently from experiment, with an assumed 50\% uncertainty. In the main text, we use the experiment to extract accurate values for the BW corrections.

\clearpage

\begin{table*}
\centering
\caption{Reduced matrix elements for the magnetic dipole (M1) and electric quadrupole (E2) transitions between the lowest few states of \ce{Ba+}.
The final column is the all-orders correlation potential, including scaling, Breit, QED, structure radiation and normalization corrections, except in the case of the non-relativistically forbidden M1 transitions, in which case it corresponds to the third-order method (see text). 
The signs are relative to the Hartree-Fock value, and numbers in square brackets refer to powers of 10.}
\label{tab:M1E2-ba}
\begin{ruledtabular}
\begin{tabular}{llllllllll}
    & &\multicolumn{4}{c}{M1 $(\mu_B)$} &  \multicolumn{4}{c}{E2 $(ea_0^2)$} \\
    \cline{3-6}\cline{7-10}
    \multicolumn{2}{c}{\ce{Ba+}}&
    \multicolumn{1}{c}{RHF} &
    \multicolumn{1}{c}{RPA} &
    \multicolumn{1}{c}{$\Sigma^{(\infty)}$} &
    \multicolumn{1}{c}{Final} &
    \multicolumn{1}{c}{RHF} &
    \multicolumn{1}{c}{RPA} &
    \multicolumn{1}{c}{$\Sigma^{(\infty)}$} &
    \multicolumn{1}{c}{Final} \\
    \hline
    \smallspace
    6$s$ & 7$s$ & 4.02\,[-5] & -14.16\,[-5] &  & -7(4)\,[-5] &  &  &  &  \\
    6$s$ & 8$s$ & 2.61\,[-5] & -8.76\,[-5] &  & -4(3)\,[-5] &  &  &  &  \\
    7$s$ & 8$s$ & 1.80\,[-5] & -3.35\,[-5] &  & -0.4(15)\,[-5] &  &  &  &  \\
    \smallspace
    6$s$ & 5$d_{3/2}$ & 0.56\,[-5] & 22.73\,[-5] &  & 16.2(33)\,[-5] & 14.763 & 14.538 & 12.594 & 12.55(13) \\
    6$s$ & 6$d_{3/2}$ & 0.72\,[-5] & 5.87\,[-5] &  & 2.3(18)\,[-5] & 18.134 & 18.108 & 16.925 & 16.81(9) \\
    6$s$ & 5$d_{5/2}$ &  &  &  &  & 18.384 & 18.141 & 15.791 & 15.71(16) \\
    6$s$ & 6$d_{5/2}$ &  &  &  &  & 21.902 & 21.863 & 20.411 & 20.28(11) \\
    7$s$ & 5$d_{3/2}$ & 0.51\,[-5] & 12.38\,[-5] &  & 8(2)\,[-5] & 6.200 & 6.277 & 4.654 & 4.66(22) \\
    7$s$ & 6$d_{3/2}$ & 0.05\,[-5] & -2.94\,[-5] &  & -0.7(11)\,[-5] & 76.978 & 76.904 & 71.033 & 70.70(11) \\
    7$s$ & 5$d_{5/2}$ &  &  &  &  & 7.936 & 8.012 & 6.050 & 6.06(28) \\
    7$s$ & 6$d_{5/2}$ &  &  &  &  & 95.043 & 94.956 & 87.810 & 87.38(15) \\
    8$s$ & 5$d_{3/2}$ & 0.35\,[-5] & 8.38\,[-5] &  & 5(2)\,[-5] & 1.727 & 1.774 & 1.494 & 1.465(47) \\
    8$s$ & 6$d_{3/2}$ & 0.13\,[-5] & 2.19\,[-5] &  & 0.7(7)\,[-5] & 39.869 & 39.908 & 35.670 & 35.55(20) \\
    8$s$ & 5$d_{5/2}$ &  &  &  &  & 2.186 & 2.230 & 1.905 & 1.874(57) \\
    8$s$ & 6$d_{5/2}$ &  &  &  &  & 50.375 & 50.420 & 45.270 & 45.10(27) \\
    \smallspace
    6$p_{1/2}$ & 7$p_{1/2}$ & 1.95\,[-5] & 2.19\,[-5] &  & 2.8(3)\,[-5] &  &  &  &  \\
    6$p_{1/2}$ & 8$p_{1/2}$ & 1.31\,[-5] & 1.47\,[-5] &  & 1.9(2)\,[-5] &  &  &  &  \\
    6$p_{1/2}$ & 6$p_{3/2}$ & 1.15275 & 1.15273 & 1.15235 & 1.15231(6) & 31.319 & 30.991 & 28.298 & 28.27(8) \\
    6$p_{1/2}$ & 7$p_{3/2}$ & 0.03638 & 0.03644 & 0.03997 & 0.0456(31) & 17.059 & 17.196 & 15.720 & 15.58(11) \\
    6$p_{1/2}$ & 8$p_{3/2}$ & 0.01828 & 0.01835 & 0.02012 & 0.0240(21) & 5.592 & 5.677 & 5.327 & 5.253(36) \\
    7$p_{1/2}$ & 8$p_{1/2}$ & 0.98\,[-5] & 1.08\,[-5] &  & 1.33(15)\,[-5] &  &  &  &  \\
    7$p_{1/2}$ & 6$p_{3/2}$ & 0.03894 & 0.03904 & 0.04299 & 0.0361(32) & 20.697 & 20.825 & 19.458 & 19.27(11) \\
    7$p_{1/2}$ & 7$p_{3/2}$ & 1.15253 & 1.15252 & 1.15216 & 1.15219(6) & 114.206 & 114.098 & 107.109 & 106.88(15) \\
    7$p_{1/2}$ & 8$p_{3/2}$ & 0.03772 & 0.03773 & 0.04057 & 0.0428(14) & 56.484 & 56.548 & 52.781 & 52.72(12) \\
    8$p_{1/2}$ & 6$p_{3/2}$ & 0.01896 & 0.01905 & 0.02090 & 0.0165(21) & 6.093 & 6.172 & 5.852 & 5.776(31) \\
    8$p_{1/2}$ & 7$p_{3/2}$ & 0.04043 & 0.04045 & 0.04371 & 0.0402(15) & 67.811 & 67.872 & 64.485 & 64.34(16) \\
    8$p_{1/2}$ & 8$p_{3/2}$ & 1.15245 & 1.15245 & 1.15211 & 1.15215(6) & 297.318 & 297.268 & 282.911 & 282.53(17) \\
    6$p_{3/2}$ & 7$p_{3/2}$ & 5.65\,[-5] & 3.54\,[-5] &  & 7.5(20)\,[-5] & 19.275 & 19.408 & 18.008 & 17.84(10) \\
    6$p_{3/2}$ & 8$p_{3/2}$ & 3.75\,[-5] & 2.38\,[-5] &  & 5.2(15)\,[-5] & 5.993 & 6.076 & 5.749 & 5.674(33) \\
    7$p_{3/2}$ & 8$p_{3/2}$ & 2.96\,[-5] & 2.17\,[-5] &  & 3.9(10)\,[-5] & 63.113 & 63.176 & 59.646 & 59.54(13) \\
    \smallspace
    5$d_{3/2}$ & 6$d_{3/2}$ & 11.82\,[-5] & 20.85\,[-5] &  & 16(3)\,[-5] & 9.843 & 9.950 & 8.214 & 8.24(22) \\
    5$d_{3/2}$ & 5$d_{5/2}$ & 1.54889 & 1.54976 & 1.54973 & 1.54928(21) & 8.091 & 7.849 & 6.750 & 6.66(10) \\
    5$d_{3/2}$ & 6$d_{5/2}$ & 0.01365 & 0.01422 & 0.01658 & 0.031(8) & 6.328 & 6.387 & 5.249 & 5.27(14) \\
    6$d_{3/2}$ & 5$d_{5/2}$ & 0.01401 & 0.01412 & 0.01663 & 0.001(8) & 6.670 & 6.733 & 5.619 & 5.64(15) \\
    6$d_{3/2}$ & 6$d_{5/2}$ & 1.54896 & 1.54902 & 1.54896 & 1.54892(2) & 49.196 & 49.113 & 45.126 & 44.90(10) \\
    5$d_{5/2}$ & 6$d_{5/2}$ & 23.26\,[-5] & 14.52\,[-5] &  & 23(4)\,[-5] & 13.105 & 13.235 & 11.005 & 11.04(29)
\end{tabular}
\end{ruledtabular}
\end{table*}

\begin{table*}
\centering
\caption{
Reduced matrix elements for the magnetic dipole (M1) and electric quadrupole (E2) transitions between the lowest few states of \ce{Ra+}.
Signs are relative to the Hartree-Fock value, and numbers in square brackets refer to powers of 10.}
\label{tab:M1E2-ra}
\begin{ruledtabular}
\begin{tabular}{llllllllll}
    & &\multicolumn{4}{c}{M1 $(\mu_B)$} &  \multicolumn{4}{c}{E2 $(ea_0^2)$} \\
    \cline{3-6}\cline{7-10}
    \multicolumn{2}{c}{\ce{Ra+}}&
    \multicolumn{1}{c}{RHF} &
    \multicolumn{1}{c}{RPA} &
    \multicolumn{1}{c}{$\Sigma^{(\infty)}$} &
    \multicolumn{1}{c}{Final} &
    \multicolumn{1}{c}{RHF} &
    \multicolumn{1}{c}{RPA} &
    \multicolumn{1}{c}{$\Sigma^{(\infty)}$} &
    \multicolumn{1}{c}{Final} \\
    \hline
    \smallspace
    7$s$ & 8$s$ & 5.14\,[-5] & -186.25\,[-5] &  & -108(40)\,[-5] &  &  &  &  \\
    7$s$ & 9$s$ & 3.35\,[-5] & -115.43\,[-5] &  & -59(30)\,[-5] &  &  &  &  \\
    8$s$ & 9$s$ & 2.17\,[-5] & -48.74\,[-5] &  & -19(15)\,[-5] &  &  &  &  \\
    \smallspace
    7$s$ & 6$d_{3/2}$ & 0.64\,[-5] & 213.77\,[-5] &  & 144(40)\,[-5] & 17.263 & 16.922 & 14.690 & 14.65(12) \\
    7$s$ & 7$d_{3/2}$ & 0.73\,[-5] & 63.53\,[-5] &  & 19(22)\,[-5] & 15.113 & 15.143 & 14.285 & 14.20(12) \\
    7$s$ & 6$d_{5/2}$ &  &  &  &  & 21.772 & 21.459 & 18.880 & 18.75(15) \\
    7$s$ & 7$d_{5/2}$ &  &  &  &  & 17.716 & 17.721 & 16.559 & 16.50(14) \\
    8$s$ & 6$d_{3/2}$ & 0.58\,[-5] & 103.63\,[-5] &  & 58(23)\,[-5] & 10.000 & 10.108 & 7.643 & 7.56(26) \\
    8$s$ & 7$d_{3/2}$ & 0.03\,[-5] & -32.02\,[-5] &  & -07(12)\,[-5] & 82.685 & 82.559 & 76.134 & 75.76(12) \\
    8$s$ & 6$d_{5/2}$ &  &  &  &  & 13.193 & 13.272 & 10.517 & 10.46(35) \\
    8$s$ & 7$d_{5/2}$ &  &  &  &  & 102.745 & 102.610 & 94.960 & 94.50(15) \\
    9$s$ & 6$d_{3/2}$ & 0.39\,[-5] & 68.02\,[-5] &  & 36(16)\,[-5] & 2.317 & 2.383 & 2.127 & 2.070(38) \\
    9$s$ & 7$d_{3/2}$ & 0.17\,[-5] & 21.43\,[-5] &  & 5(8)\,[-5] & 53.352 & 53.415 & 47.423 & 47.10(24) \\
    9$s$ & 6$d_{5/2}$ &  &  &  &  & 2.970 & 3.016 & 2.768 & 2.720(43) \\
    9$s$ & 7$d_{5/2}$ &  &  &  &  & 69.631 & 69.696 & 62.896 & 62.51(37) \\
    \smallspace
    7$p_{1/2}$ & 8$p_{1/2}$ & 2.34\,[-5] & 1.84\,[-5] &  & 3.8(10)\,[-5] &  &  &  &  \\
    7$p_{1/2}$ & 9$p_{1/2}$ & 1.59\,[-5] & 1.22\,[-5] &  & 2.6(7)\,[-5] &  &  &  &  \\
    7$p_{1/2}$ & 7$p_{3/2}$ & 1.13774 & 1.13754 & 1.13434 & 1.1340(5) & 33.268 & 32.857 & 29.774 & 29.79(10) \\
    7$p_{1/2}$ & 8$p_{3/2}$ & 0.10024 & 0.10042 & 0.10914 & 0.1140(32) & 14.453 & 14.633 & 12.926 & 12.81(15) \\
    7$p_{1/2}$ & 9$p_{3/2}$ & 0.05182 & 0.05202 & 0.05673 & 0.0604(22) & 5.196 & 5.310 & 4.830 & 4.770(63) \\
    8$p_{1/2}$ & 9$p_{1/2}$ & 1.15\,[-5] & 0.94\,[-5] &  & 1.79(40)\,[-5] &  &  &  &  \\
    8$p_{1/2}$ & 7$p_{3/2}$ & 0.12243 & 0.12287 & 0.13507 & 0.1265(38) & 25.882 & 26.026 & 24.563 & 24.33(14) \\
    8$p_{1/2}$ & 8$p_{3/2}$ & 1.13573 & 1.13566 & 1.13273 & 1.1329(5) & 119.101 & 118.965 & 111.083 & 110.84(16) \\
    8$p_{1/2}$ & 9$p_{3/2}$ & 0.10336 & 0.10336 & 0.10993 & 0.1115(18) & 47.957 & 48.039 & 43.284 & 43.64(44) \\
    9$p_{1/2}$ & 7$p_{3/2}$ & 0.05755 & 0.05796 & 0.06324 & 0.0584(22) & 6.741 & 6.828 & 6.466 & 6.368(32) \\
    9$p_{1/2}$ & 8$p_{3/2}$ & 0.12688 & 0.12698 & 0.13691 & 0.1320(23) & 82.923 & 82.995 & 79.252 & 79.37(50) \\
    9$p_{1/2}$ & 9$p_{3/2}$ & 1.13510 & 1.13507 & 1.13235 & 1.1327(5) & 306.825 & 306.763 & 291.112 & 290.1(7) \\
    7$p_{3/2}$ & 8$p_{3/2}$ & 6.12\,[-5] & -16.08\,[-5] &  & 07(11)\,[-5] & 21.256 & 21.421 & 19.878 & 19.70(12) \\
    7$p_{3/2}$ & 9$p_{3/2}$ & 4.12\,[-5] & -10.78\,[-5] &  & 6(9)\,[-5] & 6.596 & 6.699 & 6.315 & 6.238(43) \\
    8$p_{3/2}$ & 9$p_{3/2}$ & 3.18\,[-5] & -5.68\,[-5] &  & 5(6)\,[-5] & 68.017 & 68.095 & 63.823 & 64.12(51) \\
    \smallspace
    6$d_{3/2}$ & 7$d_{3/2}$ & 12.05\,[-5] & 93.79\,[-5] &  & 29(33)\,[-5] & 11.999 & 12.143 & 9.933 & 9.90(24) \\
    6$d_{3/2}$ & 6$d_{5/2}$ & 1.54780 & 1.55497 & 1.55490 & 1.5511(18) & 10.367 & 10.109 & 8.662 & 8.55(11) \\
    6$d_{3/2}$ & 7$d_{5/2}$ & 0.02889 & 0.03188 & 0.04179 & 0.055(7) & 7.537 & 7.596 & 6.107 & 6.11(16) \\
    7$d_{3/2}$ & 6$d_{5/2}$ & 0.03069 & 0.02898 & 0.04009 & 0.025(8) & 8.409 & 8.474 & 7.192 & 7.19(16) \\
    7$d_{3/2}$ & 7$d_{5/2}$ & 1.54793 & 1.54876 & 1.54803 & 1.54758(24) & 55.883 & 55.788 & 51.149 & 50.85(11) \\
    6$d_{5/2}$ & 7$d_{5/2}$ & 22.71\,[-5] & -62.79\,[-5] &  & -20(22)\,[-5] & 16.164 & 16.316 & 13.657 & 13.64(32)
\end{tabular}
\end{ruledtabular}
\end{table*}

\begin{table*}
\caption{Hyperfine A constants (MHz) for \ce{^135Ba+} and \ce{^223Ra+}, and comparison with experiment. The `Subtotal' column includes the Breit, scaling, and structure radiation corrections; `Final' further includes the QED and Bohr-Weisskopf (BW) corrections.
The uncertainties in the `Final' column show the atomic and nuclear (BW) contributions separately.
For the nuclear magnetic dipole moments, we use
$0.8381\,\mu_N$ and $0.2692\,\mu_N$, respectively~\cite{Mertzimekis:2016,*IAEAonline}.}
\begin{ruledtabular}
    \begin{tabular}{lD{.}{.}{4.2}D{.}{.}{4.2}D{.}{.}{4.2}D{.}{.}{4.2} D{.}{.}{3.2}D{.}{.}{3.2} D{(}{(}{4.7} r}
    \multicolumn{1}{c}{\ce{^135Ba+}} & \multicolumn{1}{c}{HF}& \multicolumn{1}{c}{RPA}     & \multicolumn{1}{c}{$\Sigma^{(\infty)}$}  & \multicolumn{1}{c}{Subtotal} & \multicolumn{1}{c}{$\delta$QED~\cite{Ginges:2017fle}} & \multicolumn{1}{c}{$\delta$BW\footnotemark[1]}   & \multicolumn{1}{c}{Final}     &  \multicolumn{1}{c}{Expt.}   \\
    \hline
    \smallspace
    6$s$       & 2624.51 & 3116.08 & 3718.70 & 3640.40 & -13.54   & -45.57 & 3595(13)(14)   & 3591.670...~\cite{Trapp:2000}   \\
    7$s$       & 862.65  & 1018.17 & 1106.96 & 1097.38 & -4.45    & -13.74 & 1083.6(22)(41) &             \\
    8$s$       & 395.92  & 466.19  & 493.73  & 490.78  & -2.04    & -6.14  & 484.6(12)(18)  &             \\
    6$p_{1/2}$ & 441.36  & 530.84  & 670.85  & 673.16  & -0.79    & -0.70  & 672.5(92)      & 664.6(3)~\cite{Villemoes:1993}    \\
    7$p_{1/2}$ & 172.79  & 205.43  & 240.89  & 243.80  & -0.26    & -0.25  & 243.5(24)      &             \\
    8$p_{1/2}$ & 85.88   & 101.66  & 116.26  & 117.95  & -0.12    & -0.12  & 117.8(10)      &             \\
    6$p_{3/2}$ & 64.35   & 105.90  & 131.31  & 113.53  &          & -0.18  & 113.4(88)      & 113.0(1)~\cite{Villemoes:1993}    \\
    7$p_{3/2}$ & 25.49   & 41.24   & 47.79   & 41.39   &          & -0.06  & 41.3(32)       &             \\
    8$p_{3/2}$ & 12.75   & 20.48   & 23.19   & 20.12   &          & -0.03  & 20.1(16)       &             \\
    5$d_{3/2}$ & 114.80  & 134.21  & 164.43  & 190.43  &          & -0.82  & 190(18)        & 169.5892(9)~\cite{Silverans:1986} \\
    6$d_{3/2}$ & 24.85   & 32.89   & 34.27   & 36.57   &          & -0.07  & 36.5(14)       &             \\
    5$d_{5/2}$ & 46.06   & -50.92  & -48.37  & -14.72  &          & 0.83   & -14(17)        & -10.735(2)~\cite{Silverans:1986}  \\
    6$d_{5/2}$ & 10.09   & -0.93   & 2.42    & 6.74    &          & 0.07   & 6.8(21)        &             \\
    \hline\\
    \multicolumn{1}{c}{\ce{^223Ra+}} & \multicolumn{1}{c}{HF}& \multicolumn{1}{c}{RPA}     & \multicolumn{1}{c}{$\Sigma^{(\infty)}$}  & \multicolumn{1}{c}{Subtotal} & \multicolumn{1}{c}{$\delta$QED~\cite{Ginges:2017fle}} & \multicolumn{1}{c}{$\delta$BW}  & \multicolumn{1}{c}{Final}     &  \multicolumn{1}{c}{Expt.}   \\
    \hline
    \smallspace
    7$s$       & 2674.77 & 3135.83 & 3653.68 & 3563.73 & -19.45   & -139.87 & 3424(29)(42)  & 3404.0(19)~\cite{Neu:1988wt} \\
    & & & & & & &  & 3398.3(29)~\cite{Wendt:1986ru} \\
    8$s$       & 833.22  & 971.79  & 1031.13 & 1019.70 & -6.06    & -40.03  & 980(11)(12)   &             \\
    9$s$       & 378.36  & 440.33  & 455.89  & 452.06  & -2.75    & -17.75  & 434.3(51)(53) &             \\
    7$p_{1/2}$ & 446.61  & 532.88  & 685.42  & 684.00  & -1.14    & -8.89   & 675(10)(03)    & 667.1(21)~\cite{Wendt:1986ru}  \\
    8$p_{1/2}$ & 173.21  & 204.33  & 239.72  & 242.45  & -0.36    & -3.17   & 239.3(26)(10) &             \\
    9$p_{1/2}$ & 85.92   & 100.93  & 115.18  & 117.39  & -0.16    & -1.53   & 115.9(12)(05)  &             \\
    7$p_{3/2}$ & 33.76   & 56.14   & 70.26   & 58.01   &          & -0.28   & 57.7(60)      & 56.5(8)~\cite{Neu:1988wt}     \\
    8$p_{3/2}$ & 13.60   & 22.60   & 26.46   & 21.48   &          & -0.10   & 21.4(25)      &             \\
    9$p_{3/2}$ & 6.87    & 11.42   & 13.01   & 10.57   &          & -0.05   & 10.5(13)      &             \\
    6$d_{3/2}$ & 52.68   & 46.30   & 60.17   & 97.69   &          & -2.37   & 95(20)        &             \\
    7$d_{3/2}$ & 13.32   & 15.73   & 17.07   & 23.18   &          & -0.37   & 22.8(30)      &             \\
    6$d_{5/2}$ & 19.15   & -48.28  & -52.61  & -31.74  &          & 2.44    & -29(10)       &             \\
    7$d_{5/2}$ & 4.87    & -7.00   & -5.32   & -1.71   &          & 0.36    & -1.4(17)      &             \\
    \end{tabular}
\end{ruledtabular}
\label{tab:hfs_BaRa+}
\footnotemark[1]{Note that we use the simple single particle model (see, e.g., Refs.~\cite{Volotka:2008.78.062507,Roberts:2020vef}) to estimate the BW effect independently from experiment, with an assumed 50\% uncertainty. In the main text, we use the experiment to extract accurate values for the BW corrections.}
\end{table*}

\twocolumngrid
\clearpage
\bibliography{references}

\end{document}